\documentclass[
    superscriptaddress,
    reprint,
    showpacs,
    amsmath,
    amssymb,
    aps,
    prb,
    floatfix
]{revtex4-1}
\usepackage{bbold}
\usepackage{float}
\usepackage{graphicx}
\usepackage{dcolumn}
\usepackage{bm}
\usepackage{braket}
\usepackage{subfigure}
\usepackage{hyperref}
\usepackage{dsfont}
\hypersetup{
    colorlinks=true,
    linkcolor=blue,
    citecolor=blue,
    urlcolor=blue
}

\newcommand{\B}{\scriptscriptstyle{\text{B}}}

\newcommand{\bl}{\boldsymbol{l}}

\newcommand{\bq}{\boldsymbol{q}}
\newcommand{\bM}{\boldsymbol{M}}
\newcommand{\bPi}{\boldsymbol{\Pi}}
\newcommand{\bG}{\boldsymbol{G}}
\newcommand{\bR}{\boldsymbol{R}}
\newcommand{\bPhi}{\boldsymbol{\Phi}}

\newcommand{\bLambda}{\boldsymbol{\Lambda}}
\newcommand{\ssf}{\scriptscriptstyle{(4)}}
\newcommand{\sst}{\scriptscriptstyle{(3)}}
\newcommand{\Rcal}{\mathcal{R}}
\newcommand{\Hcal}{\mathcal{H}}
\newcommand{\bRcal}{\boldsymbol{\Rcal}}
\newcommand{\Rcalhs}{\Rcal_{\scriptscriptstyle{\text{hs}}}}
\newcommand{\bRcalhs}{\bRcal_{\scriptscriptstyle{\text{hs}}}}
\newcommand{\avg}[1]{\Braket{#1}}
\newcommand{\Bavg}[1]{\Bigl\langle #1 \Bigr\rangle}
\newcommand{\Tr}{\text{Tr}}
\renewcommand{\Im}{\text{Im}}
\renewcommand{\Re}{\text{Re}}

\begin{document}
\title{
Anharmonic phonon spectra of PbTe and SnTe in the self-consistent
harmonic approximation.
}
\author{Guilherme A. S. Ribeiro}
\affiliation{Institut de min\'eralogie, de physique des mat\'eriaux et de
cosmochimie (IMPMC), Universit\'e Pierre et Marie Curie (Paris VI), CNRS UMR
7590, IRD UMR 206, Case 115, 4 place Jussieu, 75252 Paris Cedex 05, France}
\affiliation{The Capes Foundation, Ministry of Education of Brazil. Cx.
postal 250, Brasília DF 70.040-020, Brazil}
\author{Lorenzo Paulatto}
\affiliation{Institut de min\'eralogie, de physique des mat\'eriaux et de
cosmochimie (IMPMC), Universit\'e Pierre et Marie Curie (Paris VI), CNRS UMR
7590, IRD UMR 206, Case 115, 4 place Jussieu, 75252 Paris Cedex 05, France}
\author{Raffaello Bianco}
\affiliation{ Dipartimento di Fisica, Universit\`a di Roma La Sapienza,
Piazzale Aldo Moro 5, I-00185 Roma, Italy}
\affiliation{ Graphene Labs, Fondazione Istituto Italiano di Tecnologia, Via Morego, I-16163 Genova, Italy}
\author{Ion Errea}
\affiliation{Fisika Aplikatua 1 Saila, Bilboko Ingeniaritza Eskola, University of the Basque Country (UPV/EHU),
Rafael Moreno ``Pitxitxi'' Pasealekua 3, 48013 Bilbao, Basque Country, Spain}
\affiliation{Donostia International Physics Center (DIPC),
            Manuel de Lardizabal pasealekua 4, 20018 Donostia-San Sebasti\'an,
            Basque Country, Spain}
\author{Francesco Mauri}
\affiliation{ Dipartimento di Fisica, Universit\`a di Roma La Sapienza,
Piazzale Aldo Moro 5, I-00185 Roma, Italy}
\affiliation{ Graphene Labs, Fondazione Istituto Italiano di Tecnologia, Via Morego, I-16163 Genova, Italy}

\author{Matteo Calandra}
\affiliation{Institut de min\'eralogie, de physique des mat\'eriaux et de
cosmochimie (IMPMC), Universit\'e Pierre et Marie Curie (Paris VI), CNRS UMR
7590, IRD UMR 206, Case 115, 4 place Jussieu, 75252 Paris Cedex 05, France}

\date{\today}
\begin{abstract}
At room temperature, PbTe and SnTe are efficient thermoelectrics with a cubic structure. At low temperature, SnTe undergoes a ferroelectric transition with a critical temperature 
strongly dependent on the hole concentration, while PbTe is an incipient ferroelectric.
By using the stochastic self-consistent harmonic approximation, we investigate the anharmonic phonon spectra  and  the occurrence of a ferroelectric transition in both systems.
We find that vibrational spectra strongly depends on the approximation used for the exchange-correlation kernel in density functional theory.
If gradient corrections and the theoretical volume are employed, then  the calculation of the free energy Hessian leads to phonon spectra in good agreement with experimental data for both systems. 
In PbTe, we reproduce the transverse optical mode phonon satellite detected in inelastic neutron scattering and the crossing between the transverse optical and
the longitudinal acoustic modes along the $\Gamma$X direction. In the case of SnTe,  we describe the occurrence of a ferroelectric transition from the high temperature  Fm$\overline{3}$m 
structure to the low temperature R3m one.
\end{abstract}

\maketitle

\section{\label{sec:intro}Introduction}

Thermoelectric materials are appealing for their capability of converting heat into electric power, and vice versa\cite{Golds-introthermo,Fundamentals-Behnia}. Used in conjunction with clean sources of energy, such as solar radiation, such devices may be an alternative solution for the increasing global energy demand, and other global issues such as global warming\cite{ZHENG2014486,HAMIDELSHEIKH2014337,recent-thermo-Chen,Fundamentals-Behnia}.
The development of efficient thermoelectric devices is linked to a dimensionless quantity called figure of merit, given by:
\begin{equation}
ZT = \frac{S \sigma T}{k} \,,
\end{equation}\label{eq:figure_merit}
$S$ is the Seebeck coefficient, $T$ the temperature, and $\sigma$ and $k$ are the electronic and thermal conductivities, respectively. The higher is the figure of merit, higher is the efficiency of a thermoelectric device.

In these systems anharmonic effects play an important role as they decrease the thermal conductivity via phonon-phonon scattering and, consequently, increase $ZT$. In addition, some thermoelectric compounds undergo displacive second order phase transitions driven by soft modes. Close to the second order phase transition the phonon frequencies become very soft and the potential felt by the ions strongly anharmonic. Hence, a detailed non-perturbative treatment of anharmonicity is crucial to understand the lattice-dynamics and
thermoelectric efficiency of materials. From a theoretical point of view, several approximations\cite{Hellman-TDEP,Scaild-main,PhysRevLett.111.177002,PhysRevB.89.064302} have been developed in the last years to tackle this problem and it is finally becoming possible to describe anharmonic effects beyond the perturbative regime \cite{PhysRevB.59.6182,PhysRevB.65.245402}.

Among thermoelectrics, PbTe and SnTe have drawn attention due to some of their interesting properties\cite{PhysRevLett.17.753,0022-3719-2-11-304,Cochram-Lead-1966,Delaire-PbTe-nature,PhysRevB.90.214303,PhysRevB.95.144101}. Both systems have  high figures of merit, turning them into effective thermoelectrics. Furthermore, at room temperature they have simple NaCl-like structures. The simplicity of their structure and the importance of anharmonic effects to describe their lattice-dynamical properties make them an ideal playground to validate non-perturbative theoretical approaches to
the anharmonic problem. 

Beside its thermoelectric properties, SnTe displays an intriguing ferroelectric transition at low temperatures. This transition 
towards a rhombohedral structure occurs when the transverse optical (TO) modes at the Brillouin zone (BZ) center softens during cooling. Past and recent experiments measured different transition temperatures, ranging from $0\, K$ to values around $120\, K$.
Such a variation in T$_c$ is due to the change in the intrinsic doping \cite{Koba-doping}. On the other hand, PbTe does
not undergo a  phase transition at low temperature, although it has an incipient ferroelectric nature. Moreover, recent INS experiments reported that PbTe exhibits a phonon satellite peak close to zone center, a clear fingerprint of
strong anharmonicity\cite{Delaire-PbTe-nature}.

From the theoretical point of view, both systems have been studied in
the past by using ab-initio calculations, as well as molecular
dynamics based methods
\cite{PhysRevB.91.214310,PhysRevLett.112.175501,PhysRevB.32.2302,PhysRevB.85.155203,PhysRevB.95.144101}. The
majority of them use an non-perturbative approach since previous
calculations based on perturbation theory report structural
instabilities that are not present in the measurements
\cite{Hellman-TDEP,PhysRevB.91.214310}.  Some of those methods methods
give good agreement with the experiment, specially for PbTe. However,
the results may be affected by a conjunction of factors, such as the
volume and the exchange correlation functional used in the
calculations\cite{Subedi-PbTe}. Moreover
these methods do not include the quantum nature of the ions that could
be relevant at low temperature.

In this work we present the anharmonic phonon spectra of PbTe and SnTe as function of temperature using the stochastic self-consistent harmonic approximation (SSCHA)\cite{PhysRevB.89.064302,PhysRevLett.111.177002,Raffaello-Paper}. We apply the method to both systems
and determine the magnitude of anharmonic effects on vibrational spectra and on the ferroelectric transition.   

This paper is presented in the following order: First, in section
\ref{sec:theory} we introduce the theoretical background and
methodology applied in our calculations. Then we introduce the
stochastic self-consistent harmonic approximation
(SSCHA)\cite{PhysRevB.89.064302,PhysRevLett.111.177002,Raffaello-Paper}
that includes both thermal and quantum fluctuations of the ions,
and, in this framework, the evaluation of the free energy Hessian \cite{Raffaello-Paper}. Section \ref{sec:methods} presents the parameters used in our ab-initio calculations. The main results concerning the harmonic and anharmonic phonon dispersions, and comparison between our calculations and experimental data are described in section \ref{sec:results}.     

\section{\label{sec:theory}Theory}

We study the lattice dynamic of PbTe and SnTe in the Born-Oppenheimer(BO) approximation, thus 
we consider the quantum hamiltonian for the atoms defined by
BO potential energy $V(\bR)$. With $\bR$ we are denoting in component-free notation
the quantity $R^{\alpha s}(\bl)$, which is a collective coordinate
that completely specifies the atomic configuration of the crystal.
The index $\alpha$ denotes the Cartesian direction,
$s$ labels the atom within the unit cell, and $\bl$ indicates the three dimensional lattice vector. 
In what follows we will also use a single composite index $a=(\alpha,s,\bl)$ to indicate Cartesian index, atom index and lattice vector together. 
Moreover, in general, we will use bold letters to indicate also other quantities in component-free notation. 

In order to take into account quantum effects and anharmonicity at nonperturbative level,
we use the stochastic self-consistent harmonic approximation (SSCHA)~\cite{PhysRevLett.111.177002,PhysRevB.89.064302,Raffaello-Paper}.
For a given temperature $T$, the method allows to find an approximate estimation for $F(\Rcal^{\alpha s}(\bl))$, 
the free energy of the crystal as a function of the average atomic position $\Rcal^{\alpha s}(\bl)$ 
(the \emph{centroids}). For a given centroid $\bRcal$, the SSCHA free energy is obtained
through an auxiliary quadratic Hamiltonian, the SSCHA Hamiltonian $\Hcal_{\bRcal}$.
In a displacive second-order phase transition, at high temperature the free energy 
has minimum in a high symmetry configuration $\bRcalhs$ but, on lowering
temperature, $\bRcalhs$ becomes a saddle point at the transition temperature $T_c$. 
Therefore, the free energy Hessian evaluated in $\bRcalhs$, 
$\left.\partial^2F/\partial\bRcal\partial\bRcal\right|_{\bRcalhs}$, at high temperature is positive 
definite but it develops one or multiple negative eigendirections at $T_c$. The SSCHA free energy Hessian
in a centroid $\bRcal$ can be computed by using the analytic formula (in component-free notation)~\cite{Raffaello-Paper}
\begin{equation}
\frac{\partial^2 F}{\partial\bRcal\partial\bRcal}
=\bPhi+\overset{\sst}{\bPhi}\bLambda(0)
\left[\mathds{1}-\overset{\ssf}{\bPhi}\bLambda(0)\right]^{-1}\overset{\sst}{\bPhi},
\label{eq:Hessian}
\end{equation}
with
\begin{gather}
\bPhi=\Bavg{\frac{\partial^2 V}{\partial \bR\partial \bR}}_{\rho_{\Hcal_{\bRcal}}},\nonumber\\
\,\nonumber\\
\overset{\sst}{\bPhi}=\avg{\frac{\partial^3 V}{\partial \bR\partial \bR\partial \bR}}_{\rho_{\Hcal_{\bRcal}}},\quad
\overset{\ssf}{\bPhi}=\avg{\frac{\partial^4 V}{\partial \bR\partial \bR\partial \bR \partial \bR}}_{\rho_{\Hcal_{\bRcal}}},
\label{eq:smeared_FC}
\end{gather}
where the averages are with respect to the density matrix of the SSCHA Hamiltonian
$\Hcal_{\bRcal}$, i.e. $\rho_{\Hcal_{\bRcal}}=e^{-\beta\Hcal_{\bRcal}}/\text{tr}\left[e^{-\beta\Hcal_{\bRcal}}\right]$, 
and $\beta=(k_bT)^{-1}$ where $k_b$ is the Boltzmann constant. 
In Eq.~\eqref{eq:Hessian} the value at $z=0$ of the 4th-order
tensor $\bLambda(z)$ is used. For a generic complex number $z$ it is defined, in components, by 
\begin{align}
&\Lambda^{abcd}(z)=-\frac{1}{2}\sum_{\mu\nu}\,F(z,\omega_\mu,\omega_\nu)\nonumber\\
&\times
\sqrt{\frac{\hbar}{2M_a\omega_{\mu}}}e_{\mu}^a\,\sqrt{\frac{\hbar}{2M_b\omega_{\nu}}}e^b_{\nu}\,
\sqrt{\frac{\hbar}{2M_c\omega_{\mu}}}e^c_{\mu}\,\sqrt{\frac{\hbar}{2M_d\omega_{\nu}}}e^d_{\nu},
\end{align}
with $M_a$ the mass of the atom $a$, $\omega^2_{\mu}$ and $e^a_{\mu}$ eigenvalues and corresponding eigenvectors of $D^{(S)}_{ab}=\Phi_{ab}/\sqrt{M_aM_b}$, respectively, and
\begin{align}
F(z,\omega_\nu,\omega_{\mu})=&\frac{2}{\hbar}\left[\frac{(\omega_{\mu}+\omega_{\nu})[1+n_{\B}(\omega_{\mu})+n_{\B}(\omega_{\nu})]}{(\omega_\mu+\omega_\nu)^2-z^2}\right.\nonumber\\
&-\left.\frac{(\omega_{\mu}-\omega_{\nu})[n_{\B}(\omega_{\mu})-n_{\B}(\omega_{\nu})]}{(\omega_\mu-\omega_\nu)^2-z^2}\right]
\end{align}
where $n_{\B}(\omega)=1/(e^{\beta\hbar\omega}-1)$ is the bosonic occupation number.
Evaluating through Eq.~\eqref{eq:Hessian} the free energy Hessian in $\bRcalhs$ and studying its spectrum as a function of temperature, we can  predict the occurrence of a displacive second-order phase transition and estimate the relative $T_c$. In particular, since we
are considering a crystal, we take advantage of lattice periodicity and we Fourier transform the free energy 
Hessian with respect to the lattice indexes.
Therefore, since there are $2$ atoms in the unit cell of PbTe and SnTe, we actually calculate the eigenvalues 
$\lambda^2_{\mu}(\bq)$ of the two dimensional $6\times 6$ square matrix
$\partial^2F/\partial\Rcal^{\alpha s}(-\bq)\partial\Rcal^{\beta
  t}(\bq)$ in different points $\bq$ of the Brillouin zone. 
The evaluation of the magnitude of the different term of
Eq.(\ref{eq:Hessian}) is discussed in appendix \ref{apdx:magnitude}.

As showed in Ref.~\onlinecite{Raffaello-Paper}, in the context of the SCHA it is possible to 
formulate an \emph{ansatz} in order to give an approximate expression to the one-phonon Green function $\bG(z)$ for the
variable $\sqrt{M_a}(R^a-\Rcalhs^a)$
\begin{equation}
\bG^{-1}(z)=z^2\mathds{1}-\bM^{-\frac{1}{2}}\bPhi\bM^{-\frac{1}{2}}-\bPi(z),\label{Eq:Gm1}
\end{equation}
where $\boldsymbol{G}^{-1}(0)=-\boldsymbol{D}^{(F)}$, $D_{ab}^{(F)}=\frac{1}{\sqrt{M_a M_b}}\frac{\partial^2  F}{\partial\bRcal_a\partial\bRcal_b}$,
and $\bPi(z)$ is the SSCHA self-energy, given by
\begin{equation}
\bPi(z)=\bM^{-\frac{1}{2}}\,\overset{\sst}{\bPhi}\bLambda(z)
\left[\mathds{1}-\overset{\ssf}{\bPhi}\bLambda(z)\right]^{-1}\overset{\sst}{\bPhi}\,\bM^{-\frac{1}{2}},
\end{equation}
where $M_{ab}=\delta_{ab}M_a$ is the mass matrix. 
As shown in appendix \ref{apdx:magnitude}, for the applications
considered
in the present paper, the static term
$\overset{\ssf}{\bPhi}\bLambda(0)$ is negligible with respect to the
identity matrix (or using the appendix \ref{apdx:magnitude} notation,
$\langle D4V \rangle$ is negligible) . Extending this approximation to
the dynamical case reduces the SSCHA
self-energy to the so-called {\it bubble} self-energy, namely
\begin{equation}
\bPi(z)\approx\boldsymbol{\Pi}^{(B)}(z)=\bM^{-\frac{1}{2}}\,\overset{\sst}{\bPhi}\bLambda(z)
\overset{\sst}{\bPhi}\,\bM^{-\frac{1}{2}},
\end{equation}
We then neglect the mixing between different phonon modes and
 assume that $\boldsymbol{\Pi}(z)$ is diagonal  
in the basis of the eigenvectors $e^{\alpha
  s}_{\mu}(\bq) $ of $\Phi_{\alpha s,\beta
  t}({\bf q})/\sqrt{M_s M_t}$ where $\Phi_{\alpha s,\beta
  t}({\bf q})$ is the Fourier transform of 
$\Phi_{\alpha s,\beta t}$.
We then define
\begin{equation}
\Pi_{\mu}(\bq,\omega)=\sum_{\substack{\alpha s, \beta t}}e^{\alpha
  s}_{\mu}(\bq)\Pi_{\alpha s \beta t}(\bq,\omega+i0^+)e^{\beta
  t}_{\mu}(\bq)
\end{equation}
and $\omega_\mu^2(\bq)$ are the eigenvalues 
of the Fourier transform of $\boldsymbol{D}^{(S)}$.
The phonon frequencies squared, $\Omega_{{\bf q}\mu}^{2} $, corrected
by the {\it bubble}
self-energy are than obtained as
\begin{equation}
\Omega_{{\bf q}\mu}^{2}=
\omega_{\mu}^2({\bf q})+\Re\Pi_{\mu}({\bf q},\omega_\mu({\bf q})) 
\label{eq:anh_phonon_freq}
\end{equation}

In studying the response of a lattice to neutron scattering we need the one-phonon
spectral function.
By using Eq. (\ref{Eq:Gm1}) for $\bG(z)$ we can
calculate the cross-section $\sigma(\omega)=-\omega\,\Tr\,\Im\,\bG(\omega+i0^+)/\pi$, whose peaks
signal the presence of  collective vibrational excitations (phonons) having certain energies, as they can be probed with inelastic
scattering experiments (here the chosen normalization factor is such that $\int d\omega\sigma(\omega)$ 
is equal to the total number of modes).
Again, we take advantage of the lattice periodicity and we Fourier transform 
the interesting quantities with respect to the lattice indices. In particular, we consider the
Fourier transform of the SSCHA self-energy, $\Pi_{\alpha s \beta t}(\bq,z)$.
Neglecting the mixing between different modes, the cross section is then given by
\begin{align}
&\sigma(\bq,\omega)=\nonumber\\
&\frac{1}{\pi}\sum_{\mu}\frac{-\omega\,\Im\Pi_{\mu}(\bq,\omega)}{(\omega^2-\omega^2_{\mu}(\bq)-\Re\Pi_{\mu}(\bq,\omega))^2+
(\Im\Pi_{\mu}(\bq,\omega))^2}.
\label{eq:sigma}
\end{align}. 

\section{\label{sec:methods}Methods}
We perform density-functional theory (DFT) calculations using the QUANTUM-ESPRESSO package\cite{0953-8984-21-39-395502} . For both systems the exchange-correlation interaction is treated with the Perdew-Burke-Ernzerhof (PBE) generalized gradient approximation \cite{PhysRevLett.77.3865}. 
To describe the interaction between electrons and ions, we use norm-conserving pseudopotentials\cite{Trou-Martin} for PbTe, and PAW\cite{PAW-ref}  pseudopotentials for SnTe. 
In both case the semi-core $4d$ states in valence are included for Te and Sn, and $5d$ for Pb. 
Electronic wave functions are expanded in a plane-wave basis with kinetic energy cutoffs of 65 Ry and 28 Ry for scalar relativistic pseudopotentials, for PbTe and SnTe respectively. 
Integrations over the Brillouin zone (BZ) are performed using a uniform grid of $8\times8\times8$ {\bf k}-points for PbTe, and a denser grid of $12\times12\times12$ {\bf k} for SnTe. Particular care 
must be taken in converging SnTe with respect to {\bf k}-points as the depth of the potential well as a function of phonon displacements is strongly dependent on the sampling. 
Previous calculations\cite{Hell-sup} carried out with smaller samplings are found to be underconverged (see appendix \ref{ap:k-point} for convergence tests).  
Born effective charges calculated via density-functional perturbation theory (DFPT) are included for PbTe calculations only. For both PbTe and SnTe we consider the high temperature rock-salt structure, and the PBE
optimized lattice parameters of 6.55 {\AA} and 6.42 {\AA} respectively. The effect of the thermal expansion is discussed in Appendix. \ref{ap:thermal}.

Harmonic phonon frequencies are calculated within the DFPT\cite{RevModPhys.73.515} as implemented in QUANTUM-ESPRESSO. We investigate
 the $2\times2\times2$, and then, the $4\times4\times4$ {\bf q}-point grids for both systems. Fourier interpolation is used to obtain the phonon dispersion along high symmetry lines.

To calculate the anharmonic renormalized phonons we use the SSCHA\cite{PhysRevB.89.064302,PhysRevLett.111.177002,Raffaello-Paper}. The trial Hamiltoniana is minimized in a supercell. This minimization precess requires the energies and forces acting on a supercell for a set of random configurations generated by the trial density matrix. Those elements have been calculated on $2\times2\times2$ and $4\times4\times4$ supercells using the same parameters for the harmonic DFPT calculations. The number of random configurations we use is of the order of one thousand. The difference between harmonic and anharmonic dynamical matrices is interpolated to a $14\times14\times14$ supercell for SnTe. For PbTe no interpolation is needed since the $4\times4\times4$ supercell is converged and adequate to describe the experimental results.         

\section{\label{sec:results}Results \& Discussion}
\subsection{\label{subsec:harmonic} Harmonic phonon dispersion}

\begin{figure}[t!]
   \includegraphics[width=1.0\linewidth]{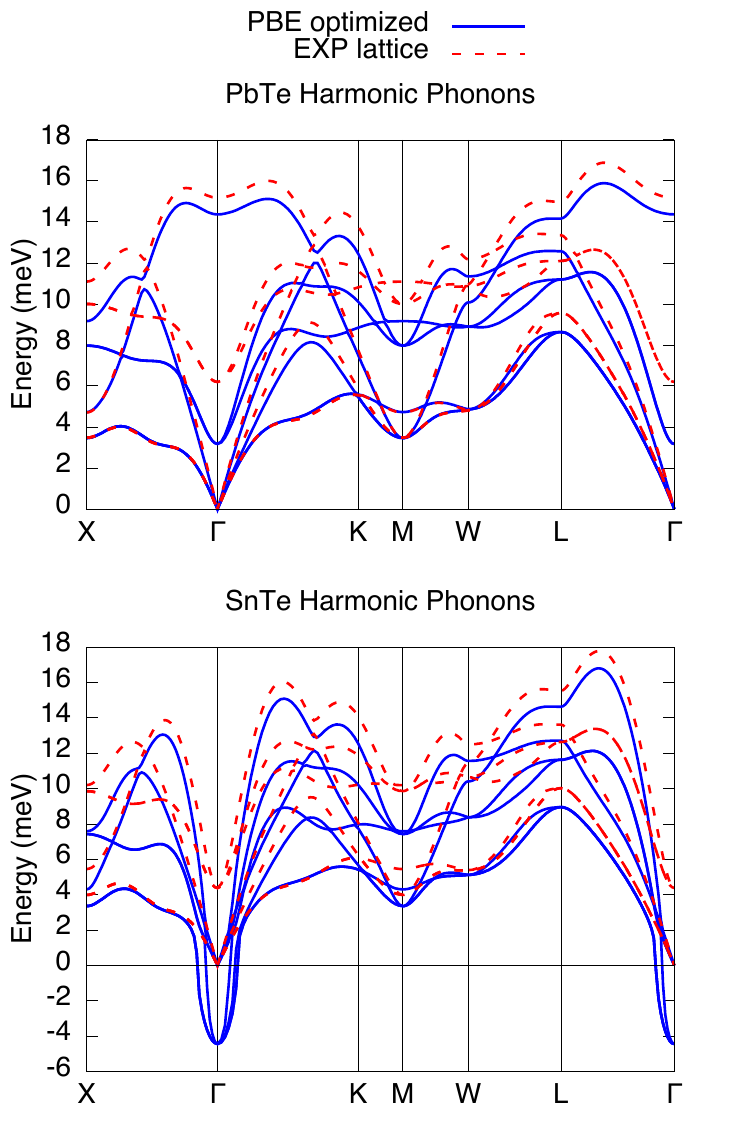}
   \caption
   {(Color online) Harmonic phonons dispersion of PbTe and SnTe for different lattice constants. For the experimental lattice parameter (red lines) both systems do not present negative frequencies. For the PBE optimized lattice parameter (blue lines), the SnTe phonon spectra has negative frequencies on $\Gamma$, indicating a structural instability. The TO modes at $\Gamma$ exhibit an strong dependence on the volume.   
}
   \label{fig:harmonicphonons}
\end{figure}

In ferroelectrics and thermoelectrics the phonon spectra strongly depends on the volume used in the calculations\cite{Subedi-PbTe}, we first investigate
the dependence of the harmonic phonon spectra on the lattice parameters. For this reason 
in Fig.(\ref{fig:harmonicphonons}) we calculate harmonic phonon dispersions within DFPT for PbTe and SnTe using (i) the zero temperature PBE theoretical 
lattice parameter and (ii) the experimental\cite{DALVEN1969141,lattice-snte} ones a$_{\rm exp}= 6.46 \AA$ and  6.32 $\AA$, respectively. In both case the experimental 
parameter is smaller then the theoretical one, as if the system were experiencing a finite pressure.
As expected, the results are strongly volume dependent.
 In the case of PbTe, the use of the experimental lattice parameter
hardens all the phonons, but the hardening is particularly large for the transverse optical (TO) mode at zone center that is shifted
from $3.17$ meV to $6.19$ meV. However in both cases, the harmonic phonon frequencies are positive and no structural instability is detected
in PbTe, in agreement with experiments. PbTe is usually referred as an incipient ferroelectric because of the softness of the TO phonon mode. It is important to underline that in the case of PbTe the experiments\cite{Cochram-Lead-1966,Delaire-PbTe-nature} find
a clear LO/TO splitting.We thus included this effect in our harmonic calculation.

At ambient pressure SnTe undergoes a phase transition in the $30-100$ K temperature range. At low temperature the crystal symmetry changes from
cubic (Fm$\overline{3}$m) to rhombohedral (R$3$m). The distortion is a displacive phase transition involving a small dimerization in the unit cell\cite{diffraction}.
The distortion is compatible with a phonon instability at zone center. Real samples of SnTe are non stoichiometric and the ferroelectric transition temperature
strongly depends on the number of holes present in the system. It is approximately 100 K for hole concentrations of the order of $1\times10^{20}$ cm$^{-3}$
and decreases to approximately 30 K for ten times larger hole concentrations. 
At these large doping no LO/TO splitting is expected, so we neglect it in the simulation.

The dependence of the harmonic calculation on volume is stronger in the case of SnTe. The ferroelectric transition (imaginary TO phonon at zone center)
is present when using the more expanded theoretical PBE volume while it disappears if the experimental volume is used. This again underlines the critical
role of the volume used in the calculation of phonon spectra in ferroelectrics and thermoelectrics\cite{PhysRevB.94.054307}.
Finally, we also investigated the role of hole doping by using the virtual crystal approximation in appendix \ref{sec:doping}
In the rest of the paper we consider the PBE optimized lattice parameter at $T= 0 \,K$ in all calculations. 
For SnTe, as we are interested in the temperatures below 100 K, we neglect the effects of the thermal expansion.  

\subsection{\label{subsec:sscha_bubble_comp} Anharmonic phonons}

\subsubsection{\label{subsec:comparison_exp} Lead telluride (PbTe)}

\begin{figure}[t!]
   \includegraphics[width=1.0\columnwidth]{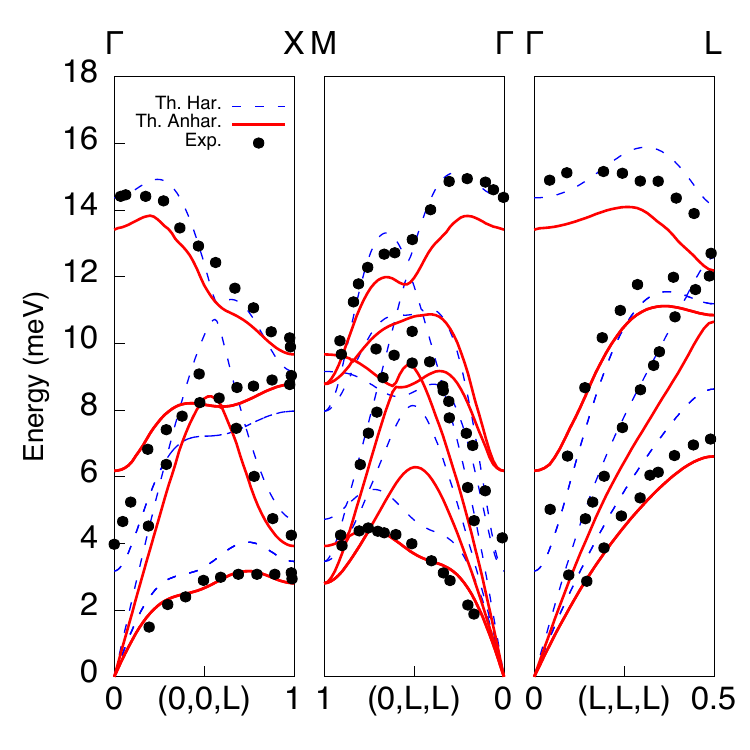}
   \caption 
   {(Color online) PbTe  harmonic (dashed lines) and anharmonic (solid lines) phonon dispersion curves at $300\,K$  compared with INS experiments\cite{Cochram-Lead-1966} at $300\,K$ (black dots). The anharmonic phonon dispersion ($\Omega_{{\bf q}\mu}$) is obtained from Eq.(\ref{eq:anh_phonon_freq}) and includes the contribution from the {\it bubble} self-energy. 
}
   \label{fig:pbte_interp}
\end{figure}

In Fig.(\ref{fig:pbte_interp}) we compare the PbTe phonon dispersion
curves $\Omega_{{\bf q}\nu}$ at $300\,K$ obtained by the SSCHA using
Eq. \ref{eq:anh_phonon_freq} to INS experimental data obtained by Cochran et al\cite{Cochram-Lead-1966}. Our calculated curves are in good agreement with experimental results. We obtain a higher value than the experiments for the TO modes at zone center is consistent to newer observations of a double peak in this region\cite{Delaire-PbTe-nature}. Previous calculations\cite{PhysRevB.85.155203} found a good agreement for the lower energy TO mode at $\Gamma$, however not obtaining a good description of the high energy phonon branches. More recent INS measurements\cite{Delaire-PbTe-nature} have suggested the presence of a strong temperature dependent phonon satellite 
close to $\Gamma$ originated from the TO mode. Furthermore, as the temperature increases, an avoided crossing between LA and TO phonon bands
along the $\Gamma$X direction is reported at $T> 300$K.  

\begin{figure}[t!]
   \includegraphics[width=1.0\linewidth]{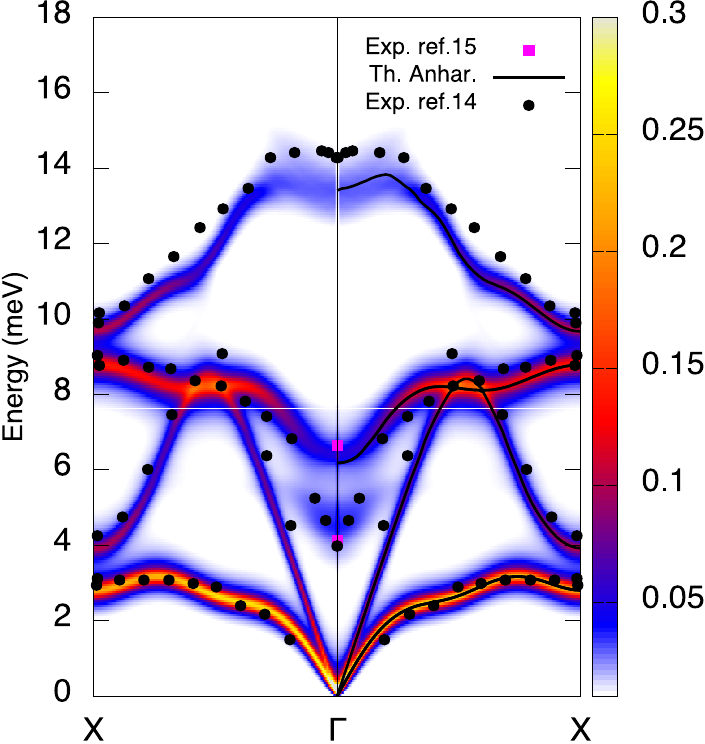}
   \caption 
   {(Color online) PbTe Spectral function at $300\,K$ calculated along the X-$\Gamma$-X path (color map). Solid lines denote the anharmonic phonon dispersion curves, black dots denote the experimental data from Ref.\onlinecite{Cochram-Lead-1966} and  pink squares denote the experimental values for the peaks at the zone center from more recent experiments described in Ref. \onlinecite{Delaire-PbTe-nature}. The color code is determined by the value of $\sigma(\bq,\omega)$ in Eq.(\ref{eq:sigma}).
}
\label{fig:tmp_dynamic}
\end{figure}

In order to determine if the SSCHA approximation can describe phonon satellites and to investigate the occurrence of the  avoided-crossing, The phonon self-energy is calculated 
performing Fourier interpolation 
over a denser $40\times40\times40$ phonon momentum
grid. Fig.(\ref{fig:tmp_dynamic}) shows our calculated anharmonic
phonon dispersion versus the neutron cross section $\sigma(\omega)$
computed using Eq. \ref{eq:sigma}
 for PbTe.

We also show, with pink dots, the Energy of the TO phonon and of its satellite as measured in INS experiments detailed in Ref.\onlinecite{Delaire-PbTe-nature}.


Both the satellite and the crossing of LA and TO bands at $300\,K$ are well described by our methodology. Moreover, the energies of the TO peaks at $\Gamma$ obtained by the SSCHA are compatible with the observed values. The presence of these features were also investigated in the literature using different methods. In particular, non-perturbative methods such as  the temperature-dependent effective potential technique (TDEP)\cite{Hellman-TDEP,PhysRevB.91.214310} were able to obtain similar results.

\subsubsection{\label{subsec:comparison_snte} Tin telluride (SnTe)}

\begin{figure}[t!]
   \includegraphics[width=1.0\linewidth]{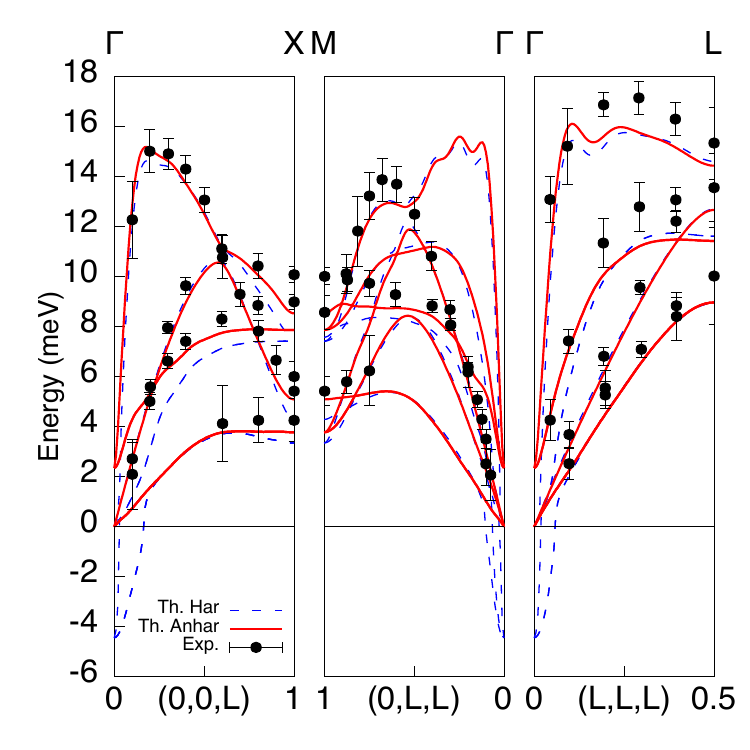}
   \caption
   {(Color online) Harmonic (dashed lines) and anharmonic (solid lines)  phonon dispersion relations of SnTe at $100\,K$ (red lines) compared with IXS experiments\cite{PhysRevB.95.144101} at $75\,K$ (black dots). The anharmonic phonon dispersion ($\Omega_{{\bf q}\mu}$) is obtained from Eq.(\ref{eq:anh_phonon_freq}) and includes the contribution form the {\it bubble} self-energy. 
}
   \label{fig:snte_interp}
\end{figure}

Calculations for  SnTe  are reported in Fig.(\ref{fig:snte_interp}) where the comparison between the phonon spectra obtained via SSCHA at $T=100\,K$ and recent IXS experiments\cite{PhysRevB.95.144101} at $T=75\,K$ is shown. Even, if this system was studied theoretically before, using methods such as the TDEP and self-consistent ab-initio
lattice dynamics (SCAILD)\cite{Scaild-main,PhysRevB.95.144101}, the
calculations focused on higher temperatures. Our calculated anharmonic
dispersion curves present the main features of the experimental data
for all investigated high symmetry directions along the BZ. Overall,
we find a good agreement with experimental data. Furthermore we also
investigate the neutron cross section $\sigma(\omega)$ throghout the
Brillouin zone but we did not detect any phonon-phonon satellite.

\begin{figure}[t!]
   \includegraphics[width=1.0\linewidth]{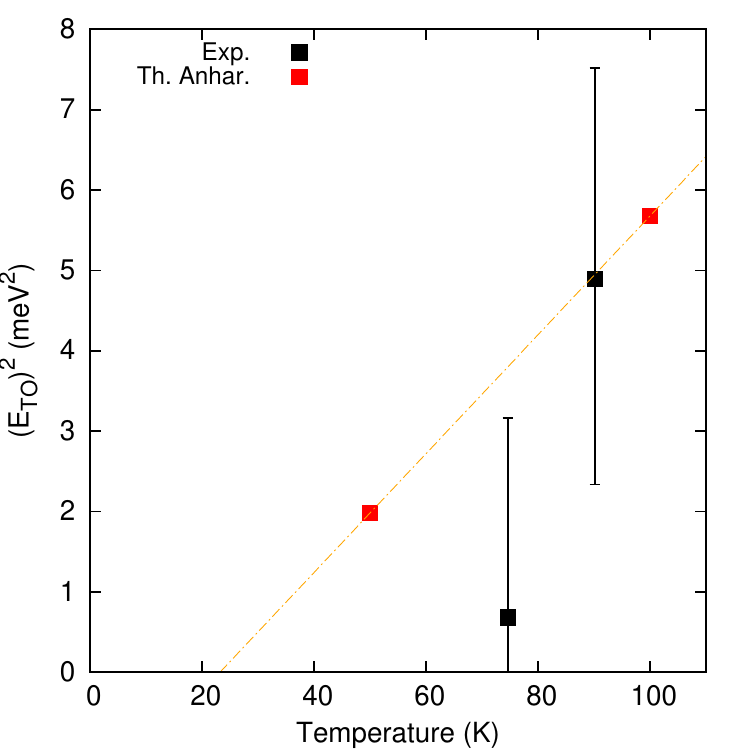}
   \caption
   {(Color online) Energy square of the TO phonons on the zone-center plotted against temperature. The red squares denote the results obtained by using Eq.(\ref{eq:anh_phonon_freq}), and black squares results form O'Neill et al\cite{PhysRevB.95.144101}. The linear extrapolation indicates that the the modes softens to zero energy towards $T_c \approx 23\, K$ for our calculations.
}
   \label{fig:tc_snte_tmp}
\end{figure}

In order to study the second order structural phase transition in SnTe we evaluate the energy squared of the TO modes at $\Gamma$ as a function of temperature $T$. 
Our data with the inclusion of anharmonicity are consistent with a ferroelectric transition at $\approx 23$ K. However, this value should be taken with care as the theoretical calculations are limited by the error in the knowledge of the exchange correlation functional that leads to a big variation in the equilibrium volume.
On the other hand, experimentally, the transition temperature of SnTe is strongly dependent on sample doping, varying from $0\,K$ to around $120\,K$ for different carriers concentrations. Fig.(\ref{fig:tc_snte_tmp}) compares our results for the energy squared of the TO mode with recent IXS\cite{PhysRevB.95.144101} experiments. 

\section{\label{sec:conlusion}Conclusions}

We applied a novel technique\cite{Raffaello-Paper} based on the stochastic self
consistent harmonic approximation capable of investigating phase
transitions via the 
calculation of the Hessian of the free energy. 
We have studied the temperature dependent anharmonic phonon spectra
of PbTe and SnTe. We found a strong dependence of vibrational
properties on the exchange-correlation functional used in the
calculation
and on the corresponding equilibrium volume.
By using the PBE\cite{PhysRevLett.77.3865} functional with the theoretical
equilibrium volume, we find very good agreement with experimental INS
spectra.
The SSCHA is not only capable of describing single particle spectra,
but also manybody features like phonon satellites are correctly 
explained.
Finally, we describe the occurrence ferroelectric transition in SnTe from the
high-T Fm$\overline{3}$m structure to the low-T R3m one. The value of the ferroelectric critical temperature is found to be strongly dependent in the voulme used in the calculations, consequently, on the exchange-correlation functional.

\begin{acknowledgments}
This work was supported by NSF Grant EFRI-143307. M.C. and F.M.  acknowledge support from the Graphene Flagship, PRACE for awarding us access to resource on Marenostrum  at BSC and the computer facilities provided by CINES, IDRIS, and CEA TGCC (Project A0010 907320). I.E. acknowledges financial support from the Spanish Ministry of Economy and Competitiveness (FIS2016-76617-P). G.R. acknowledge support from the CAPES foundation-Brazil (Grant 99999.009465/2013-08).
\end{acknowledgments}

\appendix

\section{Magnitude of $\langle D3V \rangle$ and $\langle D4V \rangle$ terms}\label{apdx:magnitude}

In order to have a better visualization of the different terms of Eq.(\ref{eq:Hessian}) one can rewrite it as:

\begin{equation}
\frac{\partial^2 F}{\partial \boldsymbol{\mathcal{R}} \partial \boldsymbol{\mathcal{R}}} = \boldsymbol{\Phi} + \overset{(3)}{\boldsymbol{\Phi}} \boldsymbol{\Lambda}(0)\overset{(3)}{\boldsymbol{\Phi}} + \overset{(3)}{\boldsymbol{\Phi}}  \boldsymbol{\Lambda}(0)\boldsymbol{\Theta}\boldsymbol{\Lambda}(0)\overset{(3)}{\boldsymbol{\Phi}}\,,
\label{eq:hessian_open}
\end{equation}
where 

\begin{equation}
\boldsymbol{\Theta} = \left[ \mathbb{1}-\overset{(4)}{\boldsymbol{\Phi}}\boldsymbol{\Lambda}(0)\right]^{-1} \overset{(4)}{\boldsymbol{\Phi}}\,. 
\end{equation}

For simplicity, we define 
\begin{eqnarray}
\left\langle D3V \right\rangle&=&\overset{(3)}{\boldsymbol{\Phi}} \boldsymbol{\Lambda}(0)\overset{(3)}{\boldsymbol{\Phi}} \\ 
\left\langle D4V \right\rangle&=&\overset{(3)}{\boldsymbol{\Phi}}  \boldsymbol{\Lambda}(0)\boldsymbol{\Theta}\boldsymbol{\Lambda}(0)\overset{(3)}{\boldsymbol{\Phi}}
\end{eqnarray}

To investigate the different terms in Eq.(\ref{eq:hessian_open}), and in particular the mutual role of $\langle D3V \rangle$ and $\langle D4V \rangle$ 
we perform SSCHA runs, on a $2\times2\times2$ supercell, for different temperatures; $300$ and $600\,K$ for PbTe, and $50$ and $100\,K$ for SnTe.
\begin{figure}[t!]
   \includegraphics[width=1.0\linewidth]{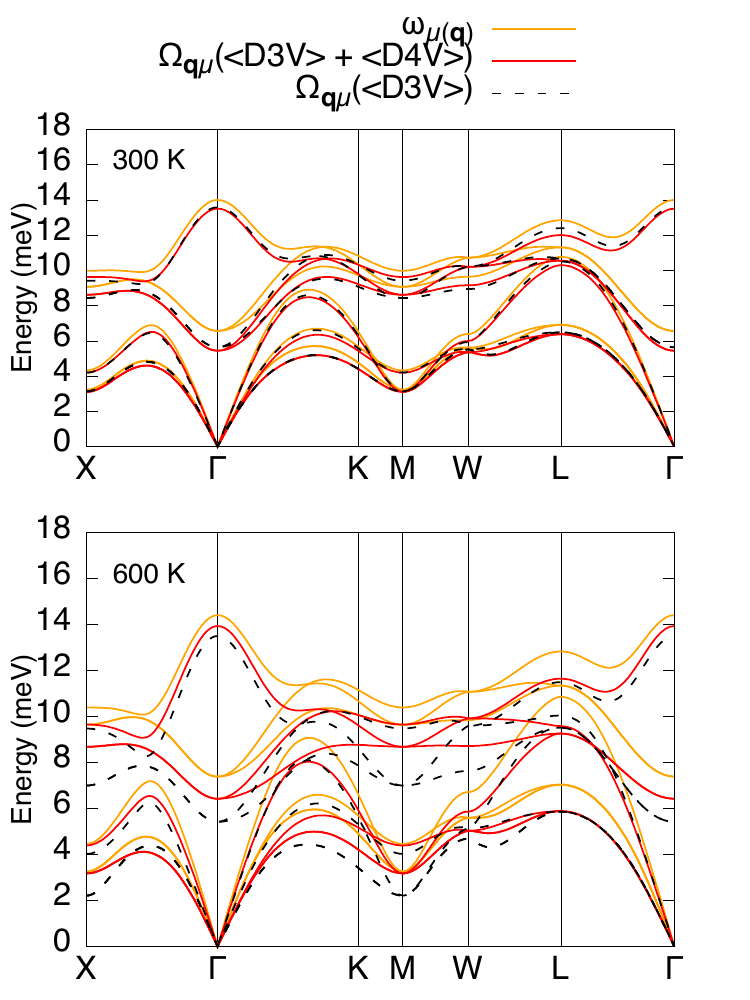}
   \caption
   {(Color online) Anharmonic phonon dispersion curves for PbTe at 300
     $K$ and 600 $K$. Orange lines denote calculations neglecting the
     {\it bubble} and superior order terms,
     ($\omega_{\mu}(\boldsymbol{q})$) ; black dashed lines represent
     the calculations including $\left\langle D3V \right\rangle$,
     while red lines includes the full expression, $\left\langle D3V
     \right\rangle + \left\langle D4V \right\rangle$,
     ($\Omega_{\mu\boldsymbol{q}}$). The phonon frequencies are
     obtained using Eq.(\ref{eq:anh_phonon_freq}) in the static limit,
     namely by using  $\Pi_\mu({\bf q},0)$.    
}
   \label{fig:snte_d3v}
\end{figure}

\begin{figure}[t!]
   \includegraphics[width=1.0\linewidth]{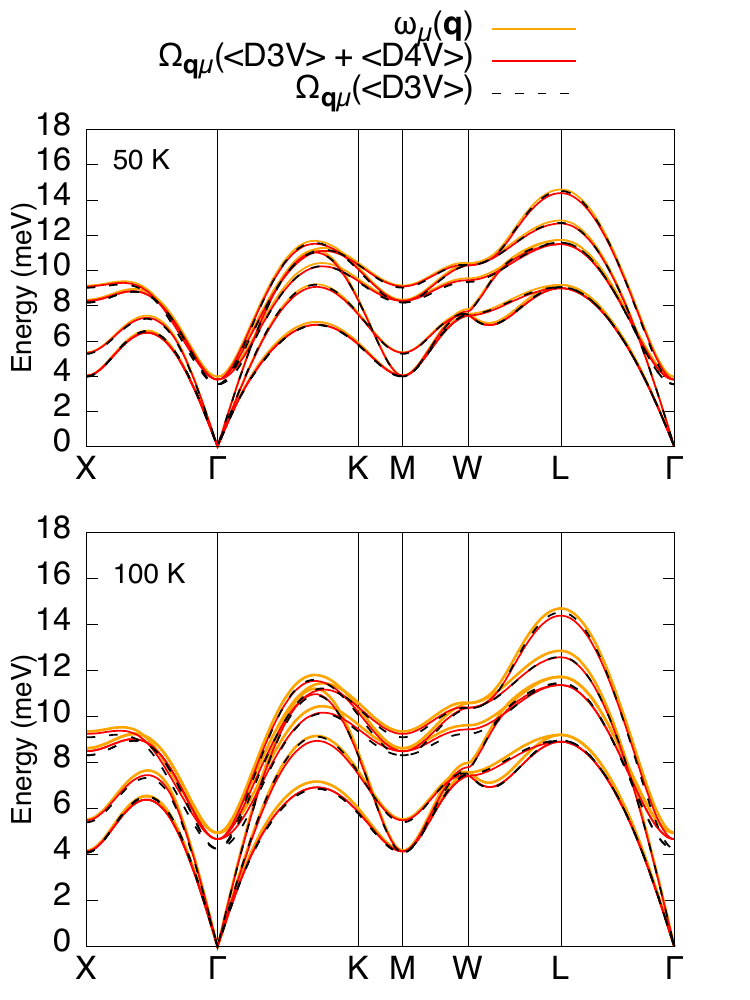}
   \caption
   {(Color online) SSCHA runs for SnTe at 50 $K$ and 100 $K$. Orange lines denote calculations neglecting the {\it bubble} and superior order terms, ($\omega_{\mu}(\boldsymbol{q})$) ; black dashed lines represent  the calculations including $\left\langle D3V \right\rangle$, while red lines includes the full expression, $\left\langle D3V \right\rangle + \left\langle D4V \right\rangle$, ($\Omega_{\mu\boldsymbol{q}}$). The phonon frequencies are obtained using Eq.(\ref{eq:anh_phonon_freq}) in the static limit,
     namely by using  $\Pi_\mu({\bf q},0)$.    . 
}
   \label{fig:pbte_d3v}
\end{figure}

Fig.(\ref{fig:snte_d3v}) and Fig.(\ref{fig:pbte_d3v}) compare the contribution of $\langle D3V \rangle$ and $\langle D4V \rangle$  to the phonon frequencies. 
Our calculations show that for PbTe the $\langle D4V \rangle$ term is negligible below 300 K while it is somewhat more relevant at $600$ K.
For SnTe in the temperature range studied in this work the $\langle D4V \rangle$ term is also negligible.
As a consequence, in these temperature regions the Hessian of the free energy is entirely determined by 
the $\boldsymbol{D}^{(S)}$ matrix and the so-called ``bubble'' term   $\langle D3V \rangle$.

This analysis justifies why we neglect the $\left\langle D4V \right\rangle$ term in the calculations for larger supercells.

Having determined the smallness of $\langle D4V \rangle$ we proceed towards larger supercell calculations. 
By using an empirical potential fitted on the SSCHA configurations we check the convergence with respect to supercell
size  (see appendix  \ref{ap:cell_size}) . We found that the use of a $4\times4\times4$ supercell leads to converged phonon frequencies. So we use this supercell to carry out our first principles calculations.

\section{k-points sampling in SnTe}\label{ap:k-point}

\begin{figure}[t!]
 \includegraphics[width=1.0\linewidth]{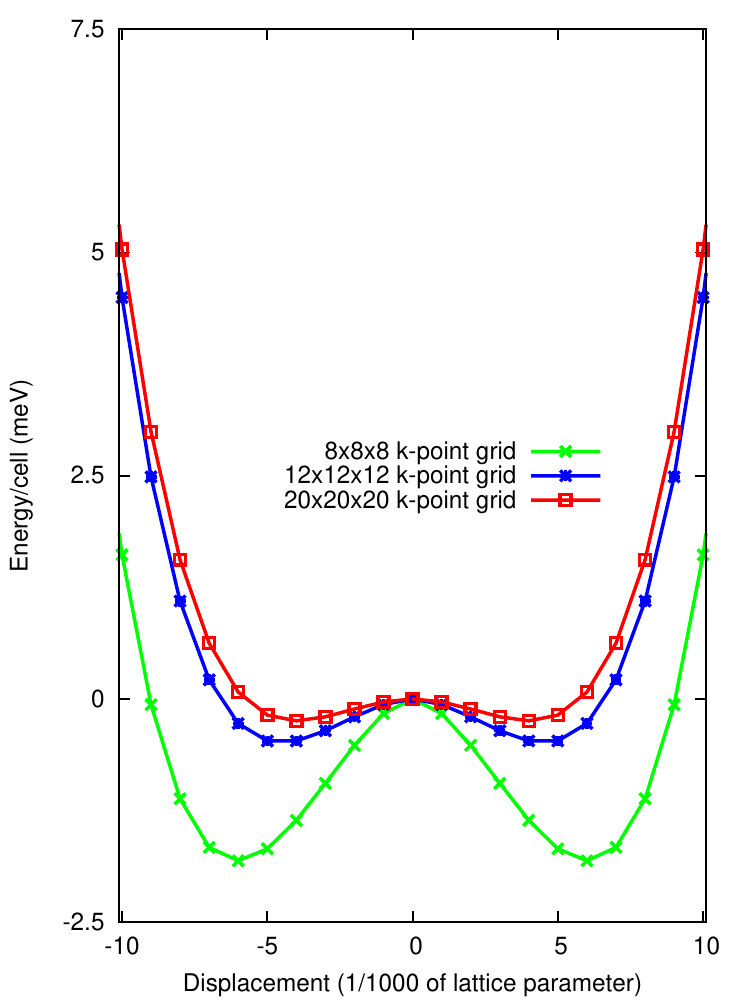}
   \caption{Frozen phonon potential for SnTe in function of the k-point grid. }
   \label{fig:SnTe_kpon}
\end{figure}

The convergence of DFT calculations for SnTe is quite tricky.  Even if
the difference on the total energy of the Fm$\overline{3}$m structure between the $8\times8\times8$ and $20\times20\times20$
k-point grids is of the order of $0.8\,$ meV/atom, upon distortion
towards the R3m structure,
the depth of the potential well differs of $1.55$ meV/cell. Thus,
using a smaller k-point grid, as done previous calculations\cite{Hell-sup}, substantially overestimates the
ferroelectric instability, as shown in Fig. \ref{fig:SnTe_kpon}. For
this reason we used a converged $12\times12\times12$ {\bf k}-point
grid in our SnTe calculations.

\section{Evaluation of Pbte's thermal expansion at 600K}\label{ap:thermal}

\begin{figure}[t!]
 \includegraphics[width=1.0\linewidth]{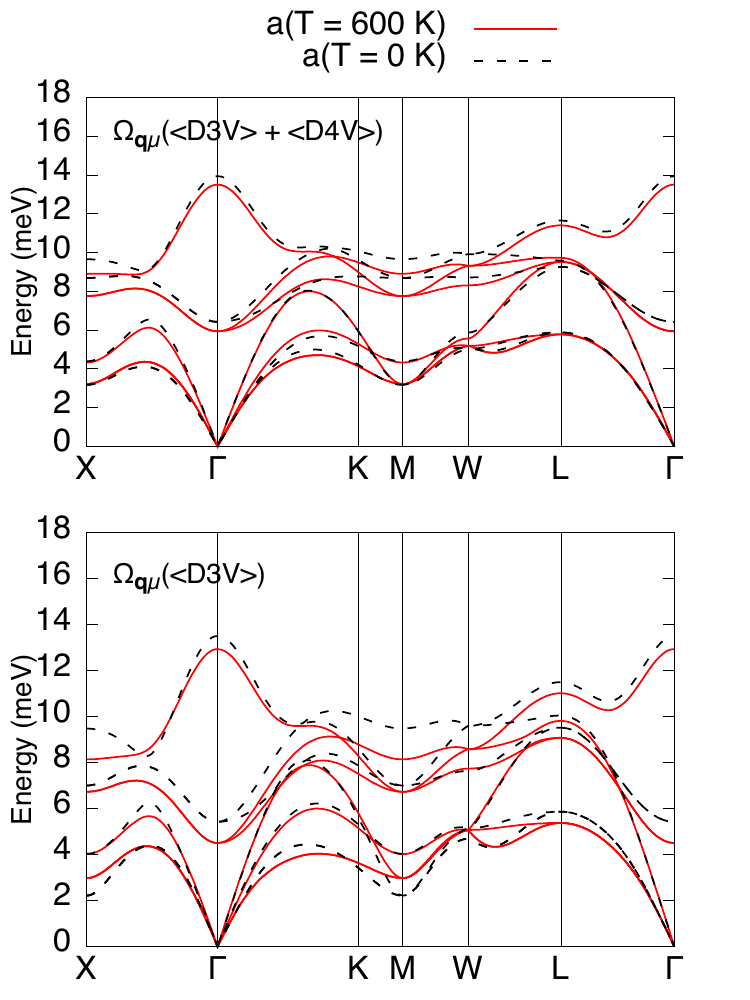}
   \caption{Anharmonic phonon spectra for the PBE at $600\,K$ lattice parameter versus the PBE at $0\,K$ case. It is noticeable that the difference is larger by including only $\left\langle D3V \right\rangle$. The phonon frequencies ($\Omega_{\mu\boldsymbol{q}}$) are obtained using Eq.(\ref{eq:anh_phonon_freq}) in the static limit,
     namely by using  $\Pi_\mu({\bf q},0)$.     }
\label{fig:therm_600Kb}
\end{figure}

To evaluate the effects of thermal expansion in our calculations for PbTe at $600\,K$, we perform several SSCHA runs for different volumes using a $2\times2\times2$ supercell. We add the vibrational free energy to the BO total energy to construct an energy vs lattice parameter curve. By finding the minimum of this curve we obtain a lattice parameter of 6.642 $\AA$ for PbTe at $600\,K$ using the PBE functional in our calculations.
With this new lattice constant, we then compute the anharmonic phonon dispersion as before. Fig.(\ref{fig:therm_600Kb}) shows the phonon spectra for the two lattice parameters (PBE $T=0\,K$ and $T=600\,K$) considering just $\left\langle D3V \right\rangle$ and the case $\left\langle D3V \right\rangle + \left\langle D4V \right\rangle$. The phonon frequencies shift towards smaller values in relation to the PBE at $0\,K$ as illustrated in Fig.(\ref{fig:pbte_d3v}). The shift is not significant for the $\left\langle D3V \right\rangle + \left\langle D4V \right\rangle$ case, whereas for the other, therms beyond the bubble gain more importance.    

\section{Born effective charges and doping}\label{sec:doping}

\begin{figure}[t!]
   \includegraphics[width=1.0\linewidth]{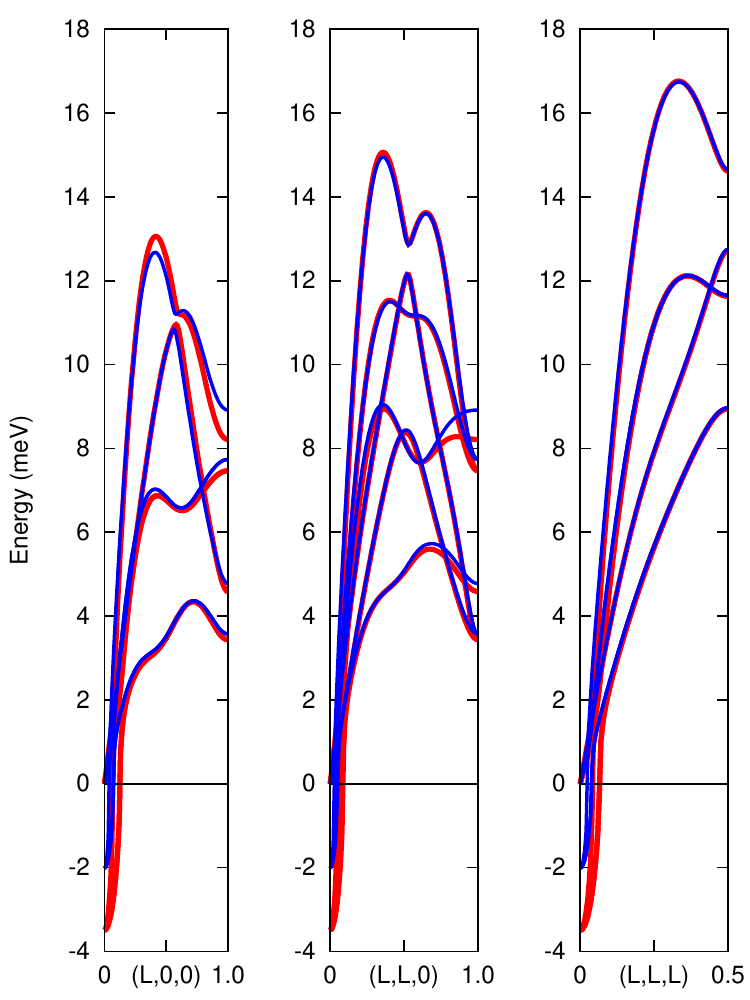}
   \caption
   {(Color online) SnTe harmonic phonon dispersion using a $4\times4\times4$ supercell: undoped (red) and doped (blue) case. Besides points very close to the zone center, doping does not change considerably the dispersion curves. 
}
   \label{fig:doping_snte}
\end{figure}

The fact that both undoped PbTe and SnTe do not exist stoichiometric raises questions on how those systems should be simulated.
For PbTe, since the typical doping in quite small and the LO-TO splitting is quite pronounced, the inclusion of the Born effective charges is a natural choice. The effective charges for PbTe are calculated via DFPT and a posteriori added to the SSCHA dynamical matrices.  On the other hand, on SnTe the effects of doping on the phonon dispersion are way more significant. Undoped SnTe should be ferroelectric, however its nature depends on the carriers concentration. Typically SnTe is heavily hole doped\cite{Koba-doping}. As hole concentration increases, the transition temperature decreases down to a point in which the system remains cubic, even at low temperatures, hence losing the ferroelectric phase.
In order to tackle this problem, we compared the harmonic phonon dispersion for doped and undoped SnTe on a $4\times4\times4$ supercell. We have not included the Born effective charges as they would be screened by doping, and we used $n_h = 3.23\times 10^{20} \, cm^{-3}$ from ref\cite{PhysRevB.95.144101} as the carrier concentration.
Fig.(\ref{fig:doping_snte}) shows that the instability remains at this doping level, and the phonon dispersion along high symmetry directions are almost unaffected within the BZ.          

\section{Empirical potential calculations}\label{ap:cell_size}

In this work, in order to investigate supercell size effects on the phonons modes, specially at the zone center, in addition to the full ab-initio calculations presented in the main text, we made use of a model potential based on the formulation developed by Marianetti et al\cite{PhysRevB.90.014308,PhysRevLett.113.105501}. The potential has the form: 

\begin{equation}
V(\boldsymbol{R}) = \frac{1}{2}\sum_{ab} \phi_{ab} u^a u^b + V_{A}^{(3)}(\boldsymbol{u}) + V_{A}^{(4)}(\boldsymbol{u})\,,
\end{equation}
 where $\boldsymbol{u} = \boldsymbol{R}-\boldsymbol{R}_{(0)}$, $\boldsymbol{R}_{(0)}$ corresponding to the equilibrium configuration on the rock-salt structure. The harmonic matrices $\phi_{ab}$ were calculated using the same parameters as for the DFPT calculations described on the main text. Anharmonic terms $V_{A}^{(3)}$ and $V_{A}^{(4)}$ are defined as:
 
\begin{equation}
V_{A}^{(3)}(\boldsymbol{u}) = p_3 \sum_{s=1}^{N_a} \sum_{\alpha = x,y,z}\Bigl[ \mathcal{A}_{s,\alpha_{+}}^3 - \mathcal{A}_{s,\alpha_{-}}^3  \Bigr]
\end{equation} 

and

\begin{align}
V_{A}^{(4)}(\boldsymbol{u}) =\, &p_4 \sum_{s=1}^{N_a} \sum_{\alpha = x,y,z}\Bigl[ \mathcal{A}_{s,\alpha_{+}}^4 + \mathcal{A}_{s,\alpha_{-}}^4  \Bigr]\nonumber\\
+\, &p_{4x} \sum_{s=1}^{N_a} \sum_{\alpha = x,y,z}\Bigl[ \mathcal{A}_{s,\alpha +}^2 \Bigl( (\mathcal{E}_{s,\alpha_{+}}^{(1)})^2 + (\mathcal{E}_{s,\alpha_{+}}^{(2)})^2 \Bigr)\nonumber \\ 
+ &\mathcal{A}_{s,\alpha_{-}}^2 \Bigl( (\mathcal{E}_{s,\alpha_{-}}^{(1)})^2 + (\mathcal{E}_{s,\alpha_{-}}^{(2)})^2 \Bigr)  \Bigr]
\end{align}

where, for example

\begin{equation}
\begin{aligned}
\mathcal{A}_{s,x_{\pm}}&=\frac{1}{\sqrt{2}}\left(u^{x_{\scriptscriptstyle{\pm}}\!(s),x}- u^{s,x}\right)\\
\mathcal{E}_{s,x_{\pm}}^{\scriptscriptstyle{(1)}}&=\frac{1}{\sqrt{2}}\left(u^{x_{\scriptscriptstyle{\pm}}\!(s),y}- u^{s,y}\right)\\
\mathcal{E}_{s,x_{\pm}}^{\scriptscriptstyle{(2)}}&=\frac{1}{\sqrt{2}}\left(u^{x_{\scriptscriptstyle{\pm}}\!(s),z}- u^{s,z}\right)\,.
\end{aligned}
\end{equation}

The variables $x_{\scriptscriptstyle{+}}(s)$ and $x_{\scriptscriptstyle{-}}(s)$ represent the nearest-neighbour of atom $s$, along the cartesian direction $+x$ and $-x$, respectively. For the other cartesian directions, $\pm y$ and $\pm z$, we generalized this notation.  The quantity $u$ is the displacement from the equilibrium position.
 
 The potentials for both systems were defined by fitting the parameters $p_3$, $p_4$, and $p_{4x}$ to ab-initio forces calculated for one thousand random atomic configurations generated in the first principles SSCHA calculation. For PbTe we used a combination of configurations generated at $300\,K$ and $600\,K$, resulting on the coefficients $p_3=2.99\,eV/(\AA)^3$, $p_4=4.17\,eV/(\AA)^4$, and $p_{4x}=-1.32\,eV/(\AA)^4$. For SnTe we used configurations generated at $100\,K$ obtaining $p_3=2.51\,eV/(\AA)^3$ and  $p_4=6.18\,eV/(\AA)^4$, in this case $p_{4x}$ was neglected since its contribution was not relevant.
 
\begin{figure}[t!]
    \includegraphics[width=1.0\linewidth]{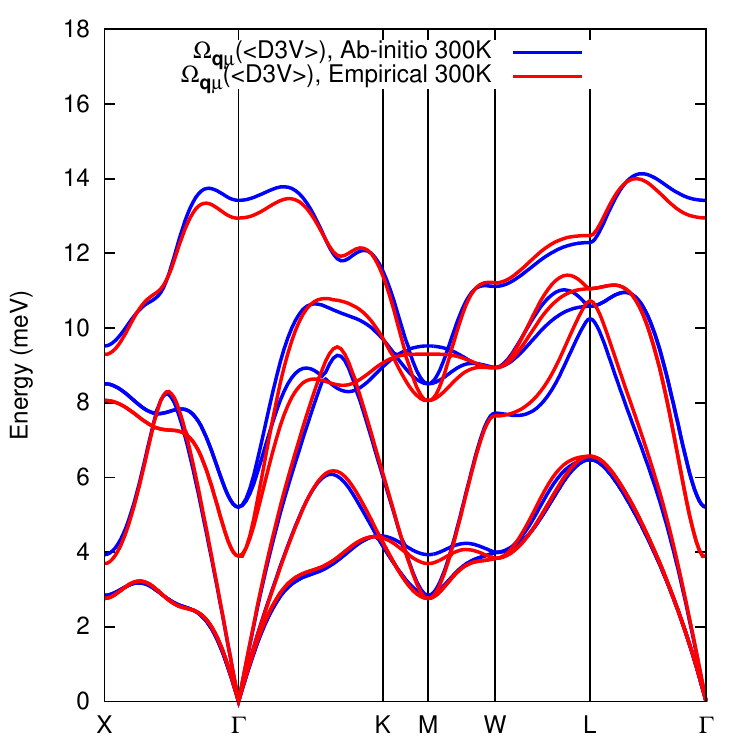}
   \caption
   {(Color online) Anharmonic PbTe phonon dispersion at $300\,K$ on a $4\times4\times4$ supercell: Ab-initio vs (blue lines) empirical potential (red lines). 
}
   \label{fig:PbTe_ab_emp}
 
 \end{figure}

\begin{figure}[b!]
    \includegraphics[width=1.0\linewidth]{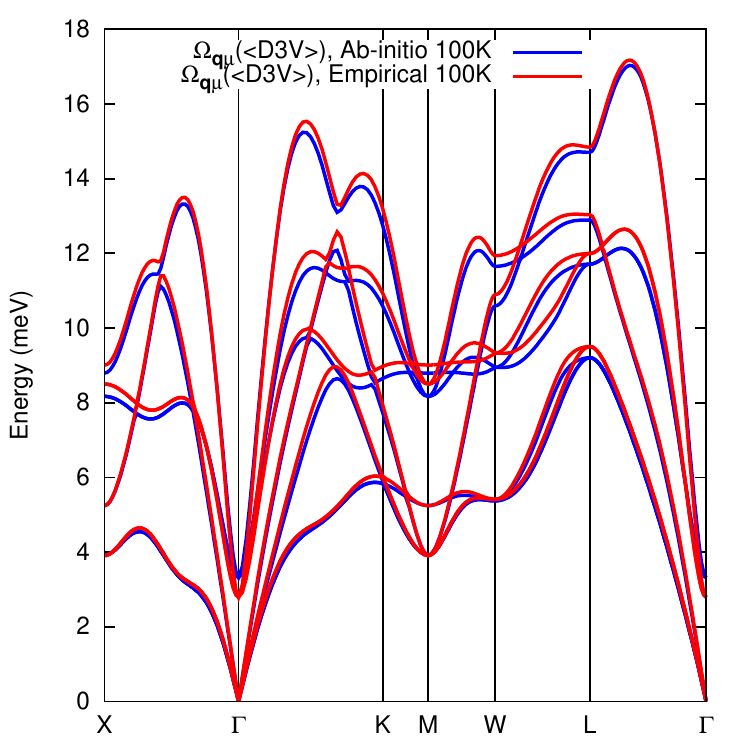}
   \caption
   {(Color online) Anharmonic SnTe phonon dispersion at $100\,K$ on a $4\times4\times4$ supercell: Ab-initio vs (blue lines) empirical potential (red lines). The phonon frequencies ($\Omega_{\mu\boldsymbol{q}}$) are obtained using Eq.(\ref{eq:anh_phonon_freq}) in the static limit,
     namely by using  $\Pi_\mu({\bf q},0)$.     
}
   \label{fig:SnTe_ab_emp}
 
 \end{figure}

 Fig.(\ref{fig:PbTe_ab_emp}) and Fig.(\ref{fig:SnTe_ab_emp}) compare the phonon dispersion on a $4\times4\times4$ supercell calculated ab-intio and using the empirical potential for PbTe ($300\,K$) and SnTe ($100\,K$), respectively.
For PbTe the larger difference is at zone center, this may be due to the fact that we used random configurations generated on a broader range of temperatures. However, this is not a problem in order to study the convergence of the TO modes using the empirical potential for different supercell sizes. For SnTe, since we generated our random configurations at $100\,K$, one may expect a better agreement. 

\begin{figure}[t!]
   \includegraphics[width=1.0\linewidth]{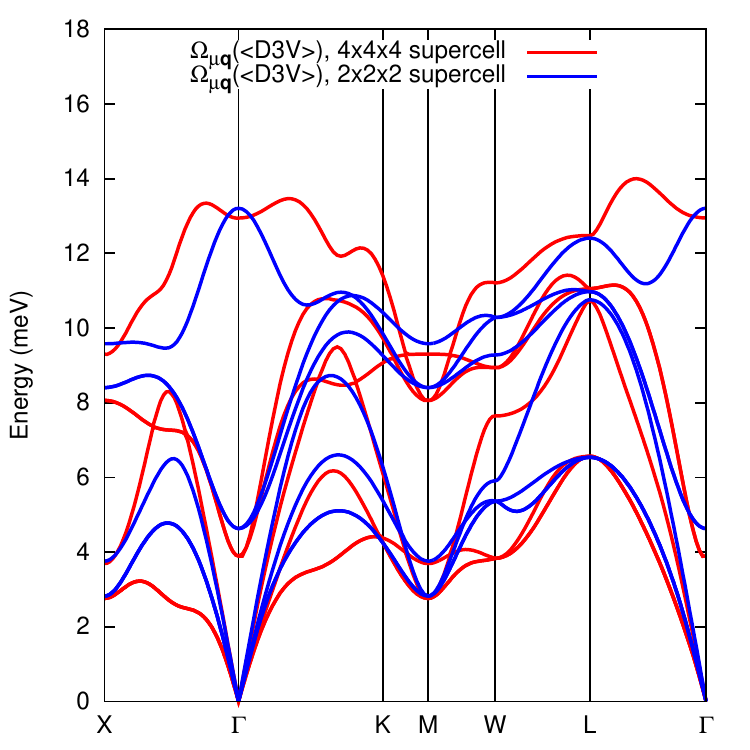}
   \caption
   {(Color online) Comparison between PbTe anharmonic phonon spectra using a $2\times2\times2$ (blue) and $4\times4\times4$ (red) supercell. Both dispersions were calculated using an empirical potential and including the bubble term. The phonon frequencies ($\Omega_{\mu\boldsymbol{q}}$) are obtained using Eq.(\ref{eq:anh_phonon_freq}) in the static limit,
     namely by using  $\Pi_\mu({\bf q},0)$.    
}
   \label{fig:PbTe_222_444}
\end{figure}

\begin{figure}[t!]
   \includegraphics[width=1.0\linewidth]{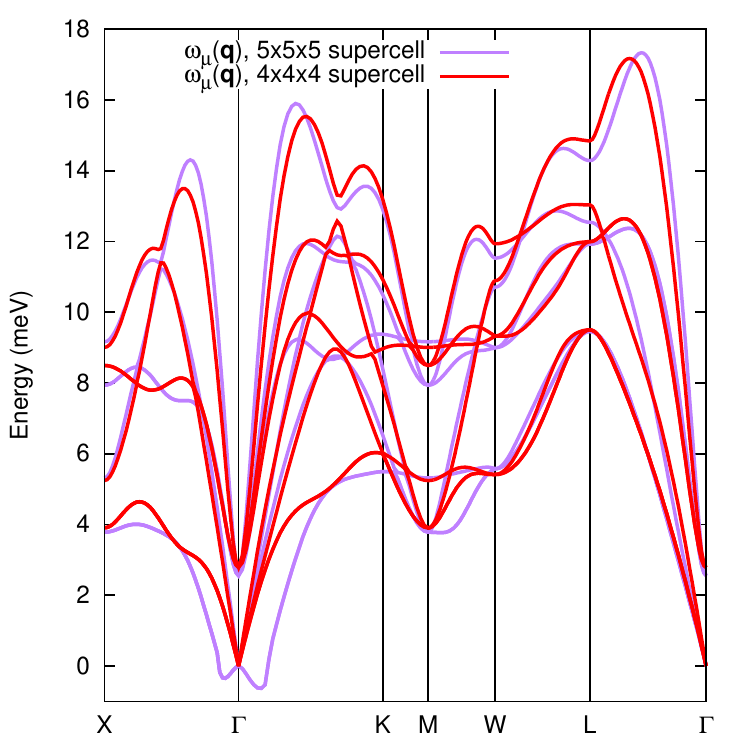}
   \caption
   {(Color online) Comparison between SnTe anharmonic phonon spectra using a $4\times4\times4$ (red) and $5\times5\times5$ (purple) supercell. In this case the {\it bubble} and higher order terms are neglected on our calculations. The phonon frequencies ($\omega_{\mu}(\boldsymbol{q})$) are obtained using Eq.(\ref{eq:anh_phonon_freq}).
}
   \label{fig:SnTe_444_555}
\end{figure}

\begin{figure}[t!]
   \includegraphics[width=1.0\linewidth]{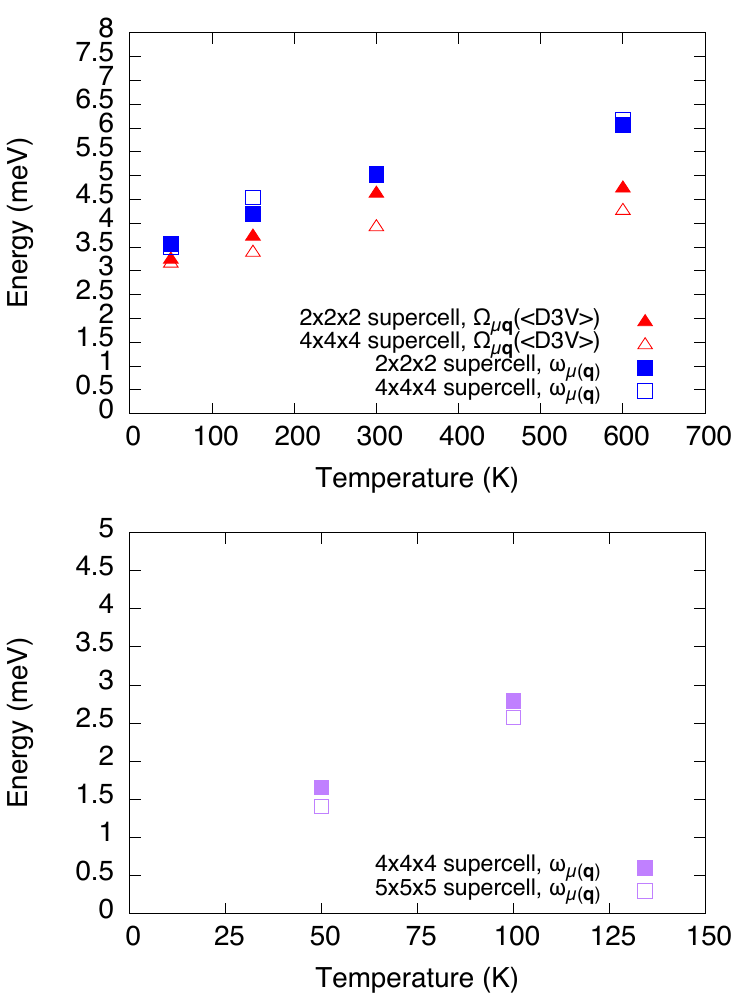}
   \caption
   {(Color online) On the TO modes in function of $T$ for PbTe (top) and SnTe (bottom) for different supercell sizes. The {\it bubble} contribution was taken on for supercells up to the $4\times4\times4$. The phonon frequencies ($\Omega_{\mu\boldsymbol{q}}$) are obtained using Eq.(\ref{eq:anh_phonon_freq}) in the static limit,
     namely by using  $\Pi_\mu({\bf q},0)$.      
}
   \label{fig:convergence_sscha}
\end{figure}

 Fig.(\ref{fig:convergence_sscha}) presents the convergence tests regarding the TO modes of PbTe and SnTe.  For the first compound, the difference between the $2\times2\times2$ and $4\times4\times4$ is small for the points which are included exactly by using the $2\times2\times2$ supercell ($X,\Gamma,L$), as shown in Fig.(\ref{fig:PbTe_222_444}). However, as stated before, we used the $4\times4\times4$ supercell in order to include more points along on the BZ and, as a consequence, to describe more accurately the phonon dispersion of PbTe. 
 As mentioned on the main text, for SnTe we considered at least a $4\times4\times4$ supercell in our ab-initio calculations, since the $2\times2\times2$ does not seem to be sufficient. In order to test the convergence we explored the model potential on the $4\times4\times4$ and also on the $5\times5\times5$ supercells. In this case, we compared just the SSCHA runs without including extra terms. 
Fig.(\ref{fig:SnTe_444_555}) shows the phonon dispersion for the $4\times4\times4$ and $5\times5\times5$ supercell. It is necessary to emphasize that the latter presents some wiggles due to the Fourier transform, so the negative energies are not physical, just an interpolation artifact.
  
As can be viewed in Fig.(\ref{fig:convergence_sscha}), the difference between the TO modes using different supercells is not significant for our purposes.

\nocite{*}
\bibliography{bibliography.bib}

\end{document}